# Anomalous Flyby in the Non-Prefered Reference Frame of the Rotating Earth

Walter Petry
Mathematisches Institut der Universitaet Duesseldorf, D-40225 Duesseldorf
E-mail: wpetry@meduse.de
petryw@uni-duesseldorf.de

Abstract: Several spacecrafts show an anomalous flyby. In a previous paper a non-prefered reference frame is studied moving uniformly relative to the prefered one. In this article the Doppler frequency residual is derived. The prefered reference frame is given by the isotropy of the CMB and the non-prefered one is the Earth. The resulting jump is much too small to explain the measured anomalous flybys of the different spacecrafts. Therefore, the transformations from the prefered frame to the non-prefered frame are replaced by the corresponding total differentials. A formula for the Doppler frequency residual is derived. It is applied to the prefered frame of the Earth and the non-prefered frame of the rotating Earth. The resulting Doppler residual depends on the direction of the velocity of the spacecraft and the position of the observer on the rotating Earth. It is similar to the experimental formula of Anderson et al. which is independent of the position of the observer.

## 1. Introduction

Anomalous Earth flybys are observed for several spacecrafts. In a previous paper [1] non-prefered frames $\Sigma$ moving relative to the prefered one $\Sigma'$ are studied. In $\Sigma'$ the pseudo-Euclidean geometry holds whereas in $\Sigma$ the light-velocity is anisotropic. Coordinate transformations from $\Sigma'$ to $\Sigma$ are stated. Several applications are given. In particular, the Michelson-Morley experiment shows no frquency shift in both frames. In the article [2] the Doppler frequency is studied where $\Sigma'$ is defined by the isotropy of the CMB and $\Sigma$ is the Earth moving relative to $\Sigma'$.

In this article the Doppler frequency formula is given in the standard form. It is similar to the Doppler formula in the prefered frame. The application of this result to calculate the jump of the spacecraft NEAR gives a much too small result. Hence, the anomalous flyby cannot be explained by these considerations. It is worth to mention that Cahill [3] has studied the same problem with anisotropic light-velocity and different considerations. He asserts that his results agree with the measured frequency jumps of all the spacecrafts but he assumes a bigger absolute value of the velocity of the non-prefered frame, i.e. of the Earth.

Furthermore, instead of the coordinate transformations of $\Sigma'$ to $\Sigma$ we consider transformations which arise out of the coordinate transformations by total differentials. A formula of the Doppler frequency residual is derived. This result is applied to the prefered frame $\Sigma'$ of the Earth and the non-prefered frame $\Sigma$ of the observer on the rotating Earth ( ground based antennas of DSN ). The Doppler residual depends on the direction of the velocity of the spacecraft, the velocity of the rotating Earth and the position of the observer. If this result gives the measured frequency jumps of all the spacecrafts the frequency jump implies not a jump of the velocity of the spacecraft. This may be the reason of the difficulty to explain flyby anomalies.

It is worth mentioning that Anderson et al. [4] are the first ones who have stated that the rotating Earth may give the anomalous flyby. An experimental fomula for the frequency jump also is stated. Our proved Doppler residual formula is similar to the experimenlal formula of Anderson et al. which doesn't depend on the position of the observer. Mbelek [5] has derived a formula of the flyby anomaly by the use of the transverse Doppler effect of special relativity with non-standard application.

Some remarks about the problem of explaining the flyby anomaly are given e.g. in [6-8].

## 2. Some Previous Results

A reference frame $\Sigma'$ is considered for which the pseudo-Euclidean geometry holds, i.e. all the results of special relativity are valid. Furthermore, a reference frame $\Sigma$ is studied which moves with constant velocity $-v' = (-v^{1'}, -v^{2'}, -v^{3'})$ relative to the frame $\Sigma'$. All the expressions in $\Sigma'$ are denoted with the same symbol as in $\Sigma$ but with a prime (see [1] ). Let $x = (x^1, x^2, x^3)$ be the Cartesian coordinats and $x^4 = ct$ then the transformations of $\Sigma'$ to $\Sigma$ and conversely are given by the formulae

$$x^{i'} = x^i \quad (i=1,2,3), \quad x^{4'} = x^4 + \left(x, \frac{v'}{c}\right)$$

$$x^i = x^{i'} \quad (i=1,2,3), \quad x^4 = x^{4'} - \left(x', \frac{v'}{c}\right). \tag{2.1}$$

Here, the symbol $(\cdot,\cdot)$ denotes the scalar product. The metric in $\Sigma$ is given by

$$(ds^2) = c^2 (d\tau)^2 = -\eta_{ij} dx^i dx^j \tag{2.2}$$

with

$$\eta_{ij} = \delta_{ij} - \frac{v^{i'}}{c}\frac{v^{j'}}{c} \qquad (i, j = 1,2,3)$$

$$\eta_{i4} = \eta_{4i} = -\frac{v^{i'}}{c} \qquad (i = 1,2,3)$$

$$\eta_{44} = -\left(1 - \left|\frac{v'}{c}\right|^2\right) \qquad (2.3)$$

where $\tau$ denotes the proper time and $|\cdot|$ is the Euclidean norm. The relations (2.1) give the connection between the three-velocity $\frac{dx'}{dt'}$ in $\Sigma'$ and the corresponding one $\frac{dx}{dt}$ in $\Sigma$:

$$\frac{dx}{dt} = \frac{dx'}{dt'}\frac{1}{1 - \left(\frac{1}{c}\frac{dx'}{dt'}, \frac{v'}{c}\right)} \qquad (2.4a)$$

$$\frac{dx'}{dt'} = \frac{dx}{dt}\frac{1}{1 + \left(\frac{1}{c}\frac{dx}{dt}, \frac{v'}{c}\right)}. \qquad (2.4b)$$

Hence, the directions of the three-velocities in $\Sigma'$ and $\Sigma$ are equal. The absolute value of the light-velocity $|v_L|$ in $\Sigma$ is given by the use of (2.4a):

$$|v_L| = \frac{c}{1 - \left|\frac{v'}{c}\right|\cos(v_L; v')}. \qquad (2.5)$$

Here, $(v_L; v')$ denotes the angle between the vectors $v_L$ and $v'$. The same result follows by the use of (2.2) and (2.3) with $d\tau = 0$.

All these results can be found in paper [1] where further applications of the theory in $\Sigma$ are given.

Let us now coinsider a plane-wave in $\Sigma$ (see article [2]). Let $k = (k_1, k_2, k_3)$ denote the wave-vector of the plane-wave and $k_4$ the relativistic fourth component. It holds

$$|k|^2 - 2\left(\frac{v'}{c}, k\right)k_4 - \left(1 - \left|\frac{v'}{c}\right|^2\right)k_4^2 = 0. \qquad (2.6)$$

Under the assumption $k_4 < 0$ relation (2.6) implies:

$$\frac{k_4}{|k|} = -\frac{1}{\left(1 - \left|\frac{v'}{c}\right|^2 \sin^2(k; v')\right)^{1/2} - \left|\frac{v'}{c}\right|\cos(k; v')} \qquad (2.7)$$

Hence, we have for the wave in $\Sigma$ by virtue of (2.5):

$$\frac{-k_4}{|k|} = \left|\frac{v_L}{c}\right| = \frac{1}{1 - \left|\frac{v'}{c}\right|\cos(v_L; v')} . \qquad (2.8)$$

The relations (2.7) and (2.8) imply

$$\left(1 - \left|\frac{v'}{c}\right|^2 \cos(k; v')\right)^{1/2} - \left|\frac{v'}{c}\right|\cos(k; v') = 1 - \left|\frac{v'}{c}\right|\cos(v_L; v'). \qquad (2.9)$$

Let us assume that a plane-wave is emitted in $\Sigma$ of an atom moving with constant velocity $w'$ (measured in $\Sigma'$). The arriving frequency $\tilde{\nu}$ of the plane-wave is by the use of $\tilde{k}_4 = \tilde{\nu}/c$ with regard to the emitted frequency $\nu$:

$$\tilde{\nu} = \gamma\nu\left(1 + \left(\frac{w'}{c}, \frac{v'}{c}\right) + \left(\left(1 - \left|\frac{v'}{c}\right|^2 \sin^2(k;v')\right)^{1/2} - \left|\frac{v'}{c}\right|\cos(k;v')\right)\left|\frac{w'}{c}\right|\cos(k;w')\right) \quad (2.10a)$$

where

$$\gamma = \left(1 - \left|\frac{w'}{c}\right|^2\right)^{-1/2}. \quad (2.10b)$$

This result is derived in article [2] where $k_4 > 0$ is assumed.

### 3. Direction of Light-Velocity

In the reference frame $\Sigma'$ the direction of the wave-vector $k'$ is equal with the direction of light-velocity. In the frame $\Sigma$ these two directions are different from one another.

It holds in $\Sigma'$ for the components $c^{i'}$ of the light-velocity:

$$c^{i'} = \alpha' k_i'$$

with a suitable constant $\alpha'$. The transfomations (2.1) of $\Sigma'$ to $\Sigma$ imply for the light-velocity in $\Sigma$ because the directions of corresponding velocities in $\Sigma'$ and $\Sigma$ are equal

$$v^i{}_L = \alpha\left(k_i - k_4 \frac{v^{i'}}{c}\right) \quad (3.1)$$

with a suitable constant $\alpha$.

It is worth to mention that the energy flux of the electro-magnetic wave is in the direction of light-velocity given by (3.1). Hence, we have to replace the direction of $k$ in (2.10a) by the direction of $v_L$. The absolute value of (3.1) gives by the use of (2.6):

$$|v_L|^2 = \alpha^2\left(|k|^2 - 2\left(k, \frac{v'}{c}\right)k_4 + \left|\frac{v'}{c}\right|^2 k_4^2\right) = \alpha^2 k_4^2.$$

Hence, it holds

$$\alpha = -|v_L|/k_4, \quad (3.2)$$

i.e. we get

$$v_L = -\frac{|v_L|}{k_4}\left(k - k_4\frac{v'}{c}\right). \quad (3.3)$$

Scalar multiplication of relation (3.3) with the vector $v'$ gives by the use of (2.7) and elementary calculations

$$\cos(k;v') = \left(\cos(v_L;v') - \left|\frac{v'}{c}\right|\right)/N \quad (3.4a)$$

with

$$N = \left(1 - 2\left|\frac{v'}{c}\right|\cos(v_L;v') + \left|\frac{v'}{c}\right|^2\right)^{1/2}. \quad (3.4b)$$

Furthermore, elementary calculations imply

$$\left(1 - \left|\frac{v'}{c}\right|^2 \sin^2(k;v')\right)^{1/2} - \left|\frac{v'}{c}\right|\cos(k;v') = N. \quad (3.5)$$

Scalar multiplication of relation (3.3) with the vector $w'$ gives by the use of (2.7) and (3.5)

$$\cos(k;w') = \left(\cos(v_L;w') - \left|\frac{v'}{c}\right|\cos(v';w')\right)/N. \quad (3.6)$$

The frequency formula (2.10a) can by the use of (3.4), (3.5) and (3.6) be rewritten in the form

$$\tilde{\nu} = \gamma \nu \left(1 + \left|\frac{w'}{c}\right| \cos(v_L; w')\right). \tag{3.7}$$

Relation (2.4b) gives

$$\left|\frac{w'}{c}\right| = \left|\frac{w}{c}\right| \frac{1}{1 + \left(\frac{w}{c}, \frac{v'}{c}\right)} \approx \left|\frac{w}{c}\right|\left(1 - \left|\frac{w}{c}\right|\left|\frac{v'}{c}\right| \cos(w, v')\right) \tag{3.8}$$

where $w$ is the velocity of the moving atom measured in the non-prefered frame $\Sigma$. Hence, the Doppler frequency residual (observed minus computed data) is

$$\frac{\Delta \nu}{\nu} \approx \left|\frac{w}{c}\right|^2 \left|\frac{v'}{c}\right| \cos(w; v'). \tag{3.9}$$

We will now apply fomula (3.9) to the flyby of NEAR. Let us assume that the prefered reference frame $\Sigma'$ is given by the isotropy of the CMB. The Earth is the non-prefered reference frame $\Sigma$ with an absolute value of the velocity $|v'| \approx 392 \frac{km}{s}$ relative to $\Sigma'$. The asymptotic value of the velocity of NEAR relative to the Earth is $|w| \approx 6.9 \frac{km}{s}$. Then, we get by the use of (3.9)

$$\frac{\Delta \nu}{\nu} \approx 6.9 \cdot 10^{-13} \cos(w; v').$$

This value is much too small to explain the jump of NEAR. Therefore, the non-prefered reference frame $\Sigma$ of the Earth moving relative to the prefered one $\Sigma'$ where the CMB is isotropic cannot explain the flyby anomaly. Cahill [3] has also studied the flyby anomaly by anisotropic light-velocity which is different from the stated here. He asserts that his results give the correct jump but he must assume a bigger value of the abolute value of the velocity of the Earth relative to the reference frame where the CMB is isotropic.

### 4. Another Form of Non-Prefered Frame

Again let us start with the prefered frame $\Sigma'$ of section 2. The transformation formulae (2.1) of $\Sigma'$ to $\Sigma$ are replaced by total differentials, i.e.

$$dx^{i\,\prime} = dx^i \quad (i=1,2,3), \quad dx^{4\,\prime} = dx^4\left(1 + \left(\frac{1}{c}\frac{dx}{dt}, \frac{v'}{c}\right)\right)$$

$$dx^i = dx^{i\,\prime} \quad (i=1,2,3), \quad dx^4 = dx^{4\,\prime}\left(1 - \left(\frac{1}{c}\frac{dx'}{dt'}, \frac{v'}{c}\right)\right). \tag{4.1}$$

The metric (2.2) is given by

$$\eta_{ij} = \delta_{ij} \qquad (i,j = 1,2,3)$$
$$\eta_{i4} = \eta_{4i} = 0 \qquad (i=1,2,3)$$
$$\eta_{44} = -\left(1 + \left(\frac{1}{c}\frac{dx}{dt}, \frac{v'}{c}\right)\right)^2 \tag{4.2}$$

where $x(t)$ is the velocity of the object moving in $\Sigma$. Hence, the coefficient $\eta_{44}$ can be time-dependent. It easily follows that the formulae (2.4) and (2.5) for the velocity of an object and of light in $\Sigma$ are also valid. It is not excluded that the velocity $v'$ in (4.1) and (4.2) is time-dependent.

Let us consider a plane-wave as before. The formulae (4.1) give for the wave in $\Sigma$

$$k_i = k_i' \quad (i=1,2,3)$$
$$k_4 = k_4'\left(1 + \left(\frac{1}{c}\frac{dx}{dt}, \frac{v'}{c}\right)\right). \tag{4.3}$$

This is contrary to the previous results in the sections 2 and 3.

Hence, the wave-vector $k$ is not changed in $\Sigma$ whereas the frequency is changed compared to the results in the prefered frame $\Sigma'$. Therefore, the direction of the light-velocity $v_L$ is in the direction of the wave-vector.

Let $v'$ be the frequency in the prefered frame $\Sigma'$. Relation (4.3) gives for the frequency in $\Sigma$:

$$v = v'\left(1 + \left(\frac{1}{c}\frac{dx}{dt}, \frac{v'}{c}\right)\right). \tag{4.4}$$

In $\Sigma'$ the well-known Doppler frequncy formula holds:

$$v' = \gamma v_0'\left(1 + \left|\frac{w'}{c}\right|\cos(v_L; w')\right) \tag{4.5a}$$

with

$$\gamma = \left(1 - \left|\frac{w'}{c}\right|^2\right)^{-1/2} \tag{4.5b}$$

Here, $w'$ is the velocity of the moving object measured in $\Sigma'$ and $v_0'$ is the frequency emitted by the same atom at rest. Then, the frequency $v$ in $\Sigma$ is by the use of (4.4) with $\frac{dx}{dt} = w$:

$$v = \gamma v_0'\left(1 + \left|\frac{w'}{c}\right|\cos(v_L; w')\right)\left(1 + \left(\frac{w}{c}, \frac{v'}{c}\right)\right) \approx \gamma v_0'\left(1 + \left|\frac{w'}{c}\right|\cos(v_L; w') + \left(\frac{w}{c}, \frac{v'}{c}\right)\right). \tag{4.6}$$

Hence, in $\Sigma$ the two-wave Doppler frequency residual is given by

$$\frac{2(v - v')}{v_0'} \approx 2\left(\frac{w}{c}, \frac{v'}{c}\right). \tag{4.7}$$

## 5. Application to Anomalous Flyby

Let us now assume that the non-prefered frame $\Sigma'$ is the Earth which is justified by the previous section since the anisotropy on the Earth caused by the isotropy of the CMB is negligible. The non-prefered frame $\Sigma$ is given by the observer on the rotating Earth, i.e. the ground based antennas of the DSN station.

Let us consider spherically symmetric coordinates on the Earth. The observer has the coordinates

$$X = \left(R\sin\left(\frac{2\pi}{T}t\right)\cos\vartheta_0, -R\cos\left(\frac{2\pi}{T}t\right)\cos\vartheta_0, R\sin\vartheta_0\right)$$

where $R$ is the radius of the Earth, T is the time of one day, and $\vartheta_0$ is the declination. The velocity of the observer is then given by

$$\frac{dX}{dt} = \frac{2\pi R}{T}\left(\cos\left(\frac{2\pi}{T}t\right)\cos\vartheta_0, \sin\left(\frac{2\pi}{T}t\right)\cos\vartheta_0, 0\right), \quad v' = -\frac{dX}{dt} \tag{5.1}$$

with fixed $t$. The velocity of the distant object (spacecraft) moving relative to the observer can be given by

$$\frac{dx}{dt} = w = |w|(\cos\varphi\cos\vartheta, \sin\varphi\cos\vartheta, \sin\vartheta) \tag{5.2}$$

where $\varphi$ and $\vartheta$ are fixed. Relation (5.1) can be rewritten:

$$v' = -\frac{2\pi R}{T}\left(\cos\left(\frac{2\pi}{T}t\right)\cos\vartheta_0, \sin\left(\frac{2\pi}{T}t\right)\cos\vartheta_0, 0\right). \tag{5.3}$$

The two-way Doppler residual (4.7) gives by the use of (5.2), (5.3) and elementary calculations

$$2\frac{v - v'}{v_0} \approx -\left|\frac{w}{c}\right|\frac{4\pi R}{cT}\cos\vartheta_0\cos\vartheta\cos\left(\frac{2\pi}{T}t - \varphi\right). \tag{5.4}$$

Light is emitted from the DSN station and moves into the direction of the spacecraft where it is reflected back to the observer. It holds by virtue of the two-way Doppler and the comparison of the equations (5.1) and (5.2)

$$tg\varphi = tg\left(\frac{2\pi}{T}t\right).$$

This implies

$$\cos\left(\frac{2\pi}{T}t - \varphi\right) = \pm 1. \tag{5.5}$$

Formula (5.4) of the two-way Doppler residual gives with the lower sign

$$2\frac{v - v'}{v_0} \approx \left|\frac{w}{c}\right|\frac{4\pi R}{cT}\cos\vartheta_0 \cos\vartheta. \tag{5.6}$$

Therefore, the frequency jump has the form

$$\frac{\Delta v}{v_0} \approx \left|\frac{w}{c}\right|\frac{4\pi R}{cT}\left(\cos\vartheta_{0a}\cos\vartheta_a - \cos\vartheta_{0b}\cos\vartheta_b\right) \tag{5.7a}$$

where the indices "$a$" and "$b$" stand for "after" and "before" the perigee. If the declinations of the positions of the observer after and before the perigee agree with one another, i.e. $\vartheta_{0a} = \vartheta_{0b} = \vartheta_0$ equation (5.7a) can be rewritten in the form

$$\frac{\Delta v}{v_0} \approx \left|\frac{w}{c}\right|\frac{4\pi R}{cT}\cos\vartheta_0 \left(\cos\vartheta_a - \cos\vartheta_b\right). \tag{5.7b}$$

The triconometric functions in formula (5.6) are non-negative by virtue of $-\frac{\pi}{2} \leq \vartheta_0, \vartheta \leq +\frac{\pi}{2}$. An observer at the poles, i.e. $\vartheta_0 = \pm\frac{\pi}{2}$ does not measure a frequency jump. Furthermore, there is no frequency jump when the spacecraft moves symmetrically about the plane of the equator $(\vartheta_a = \vartheta_b)$.

It is an open question whether the formulae (5.7) can explain the anomalous flybys of all the different spacecrafts. Indications to this are given by the experimental formula of Anderson et al. [4].

It is worth to mention that the Doppler frequency jump (5.7) does not imply a jump of the velocity of the spacecraft which is much smaller (see (2.4)).

This may be the reason of the difficulty to explain the anomalous Earth flybys.

The idee that the rotation of the Earth may explain the flyby anomaly is at first stated by Anderson et al. who have also given an experimental formula [4]. It is worth mentioning that this formula agrees with formula (5.7b) if $\cos\vartheta_0 = 1$, i.e. the observer is on the equator.

The spacecraft ROSETTA shows in 2007 and 2009 no flyby anomaly whereas the formula of Anderson et al. [4] predicts a small jump (see e.g. [6] ). It may be that formula (5.7b) implies no frequency jump if the observer has a suitable declination $\vartheta_0$ or has different declinations before and after the perigee.

Mbelek [5] also studies the rotating Earth by the use of the transverse Doppler effect of special relativity and non-standard considerations. He asserts that his result is idendical with the formula of Anderson et al..

General remarks about the problem of explaining the flyby anomaly can be found in [6-8].

Additional Remark:

The result (5.6) is a special case by virtue of the assumption (5.5). The general two-way Doppler residual is given by formula (5.4). Let us replace $\frac{2\pi}{T}t$ by the angle $\varphi_0$ when the photon arrives at the DSN-station. It holds by the comparison of (5.1) with (5.2):

$$-\cos\left(\frac{2\pi}{T}t\right) = \cos\varphi_0, \quad -\sin\left(\frac{2\pi}{T}t\right) = \sin\varphi_0$$

i.e.,we have

$$\frac{2\pi}{T}t = \varphi_0 + \pi,$$

implying

$$\cos\left(\frac{2\pi}{T}t - \varphi\right) = -\cos(\varphi - \varphi_0).$$

Formula (5.4) is now rewritten in the form

$$2\frac{v - v'}{v_0} \approx \left|\frac{w'}{c}\right|\frac{4\pi R}{cT}\cos\vartheta_0 \cos\vartheta\cos(\varphi_0 - \varphi). \tag{5.8}$$

Hence, the Doppler frequency residual (5.8) depends on the the asymptotic velocity of the spacecraft and the position $\vartheta_0, \varphi_0$ of the observer when the photon arrives. For $\varphi_0 = \varphi$ formula (5.8) gives (5.6).

References
[1] W. Petry, Astrophys.&Space Sci. **124** (1986),63
[2] W. Petry, arXiv: 0909.5150
[3] R.T. Cahill, arXiv: 0804.0039
[4] J.D.Anderson, J.K. Campbell, J.E. Ekelund and J.F. Jordan, Physical Review Letters **100** (2008), 091102
[5] J.P. Mbelek, arXiv: 0809.1888
[6] http://www.wissenschaft-online.de/artikel/1018579
[7] http://wissenschaft.de/wissenschaft/news/313590.html
[8] http://www.raumfahrer.net/news/raumfahrt/21092008194137.shtml